\documentclass[pre,aps,twocolumn,superscriptaddress,showpacs]{revtex4-1}
\usepackage{amsmath,bm,graphicx}
\usepackage{amssymb}
\usepackage{amsfonts}
\usepackage{subfigure}
\newcommand{\w}{\omega}
\renewcommand{\k}{\mathbf{k}}
\newcommand{\x}{\mathbf{x}}
\newcommand{\dd}[2]{\frac{d #1}{d #2}}

\begin{document}
\def \figwidth {\columnwidth}
\title{Percolation transition in the kinematics of nonlinear resonance broadening in Charney-Hasegawa-Mima model of Rossby wave turbulence}
\date{\today}

\author{Jamie Harris}
\affiliation{Centre for Complexity Science, University of Warwick, Gibbet Hill Road, Coventry CV4 7AL, UK}
\author{Colm Connaughton}
\affiliation{Warwick Mathematics Institute, University of Warwick, Gibbet Hill Road, Coventry CV4 7AL, UK}
\affiliation{Centre for Complexity Science, University of Warwick, Gibbet Hill Road, Coventry CV4 7AL, UK}
\author{Miguel D. Bustamante}
\affiliation{School of Mathematical Sciences, University College Dublin, Belfield, Dublin 4, Ireland}

\begin{abstract}
We study the kinematics of nonlinear resonance broadening of interacting Rossby waves as modelled by the Charney-Hasegawa-Mima equation on a biperiodic domain.  We focus on the set of wave modes which can interact quasi-resonantly at a particular level of resonance broadening and aim to characterise how the structure of this set changes as the level of resonance broadening is varied. The commonly held view that resonance broadening can be thought of as a thickening of the resonant manifold is misleading. We show that in fact the set of modes corresponding to a single quasi-resonant triad has a nontrivial structure and that its area in fact diverges for a finite degree of broadening. We also study the connectivity of the network of modes which is generated when quasi-resonant triads share common modes. This network has been argued to form the backbone for energy transfer in Rossby wave turbulence. We show that this network undergoes a percolation transition when the level of resonance broadening exceeds a
critical value. Below this critical value, the largest connected component of the quasi-resonant network contains a negligible fraction of the total number of modes in the system whereas above this critical value a finite fraction of the total number of modes in the system are contained in the largest connected component. We argue that this percolation transition should correspond to the transition to turbulence in the system.
\end{abstract}
\maketitle

\section{Introduction}
The phenomenon of dispersive wave propagation is fundamental to our understanding of a wide variety of spatially extended physical systems. In such systems,  the frequency, $\w_\k$, of each wave mode is a nonlinear function of its wave-vector, $\k$ \cite{whitham_linear_1974}. Examples include gravity-capillary waves on fluid interfaces \cite{zakharov_kolmogorov_1992}, flexural waves in thin elastic plates \cite{graff_wave_1991}, drift waves in strongly magnetized plasmas \cite{horton_quasitwodimensional_1994} and Rossby waves in planetary oceans and atmospheres \cite{pedlosky_geophysical_1987}. In this article, we will be interested in Rossby waves as modeled by the Charney-Hasegawa-Mima (CHM)
equation \cite{horton_quasitwodimensional_1994} on the $\beta$-plane:
\begin{equation}
\label{eq-CHM}
\partial_t\left(\Delta \psi - F \psi \right) + \beta \partial_x \psi +  J\left[\psi,\Delta \psi\right]= 0.
\end{equation}
This is the simplest two-dimensional model of the large scale dynamics of a shallow layer of fluid on the surface of a strongly rotating spehere. The surface of the sphere is approximated locally by a plane, $\x \in \mathbb{R}^2$, with $x$ varying in the longitudinal (meridional) direction and the $y$ varying in the latitudinal (zonal) direction. The field $\psi(\x,t)$ is the geopotential height, $\beta$ is the Coriolis parameter measuring the variation of the Coriolis force with latitude, $F$ is the inverse of the square of the deformation radius and $J\left[f,g\right]$ denotes the Jacobian of two functions, $f$ and $g$. This equation admits harmonic solutions, $\psi(\x,t) = \mathrm{Re}\left[ A_\k\, \exp(i\k\cdot\x - i\w_\k t)\right]$ with $\k \in \mathbb{R}^2$. These solutions, known as Rossby waves, have the  anisotropic dispersion relation
\begin{equation}
\label{eq-CHMdispersion}
\omega(\k) = -\frac{\beta k_x}{k^2 + F}.
\end{equation}

Since Eq.\eqref{eq-CHM} is nonlinear, modes with different wave-vectors couple together and exchange energy. If the nonlinearity is weak, one finds that this energy exchange is generally quite slow and occurs most efficiently between groups of
modes which are in {\em resonance}. For the CHM equation, such resonances involve three modes since the nonlinearity is quadratic. Four modes would be involved in the case of systems with cubic nonlinearity. Three wave-vectors $(\k_1, \k_2, \k_3)$  satisfying the resonance conditions,
\begin{equation}
\label{eq-resonance}
\left\{
\begin{aligned}
&\k_3 = \k_1 + \k_2,\\
&\omega(\k_3)-\omega(\k_1)-\omega(\k_2) =0,
\end{aligned}
\right.
\end{equation}
are referred to as a resonant triad. If one projects the spectral representation of the wave equation onto a resonant triad, one obtains a set of ordinary differential equations for the coupled time evolution of the amplitudes of the constituent modes. Such systems of equations appeared as basic models of nonlinear mode coupling in a variety of physical systems including plasma physics \cite{sagdeev_nonlinear_1969}, nonlinear optics \cite{armstrong_interactions_1962} and oceanic internal waves \cite{mccomas_resonant_1977}. An advantage of such models is that the equations of motion for a resonant triad are simple enough that explicit formulae can be obtained for both amplitudes and phases of the resonant modes \cite{craik_wave_1988, lynch2004pulsation, bustamante_effect_2009, bustamante2011resonance, harris_externally_2012}.

A disadvantage, however, is that such triads are generally not closed. Even if energy is initially mostly restricted to a single triad, other resonant modes can be generated which are not in the original triad. This process can repeat in a cascade-like fashion and result in the excitation of a large number of modes. If a large number of degrees of freedom are excited, a statistical description of energy transfer between modes is preferable. Such a description is
provided by the theory of wave turbulence \cite{nazarenko_wave_2011,zakharov_kolmogorov_1992}. This theory provides a kinetic description of energy transfer in ensembles of weakly interacting dispersive waves in which conserved quantities are redistributed along the resonant manifolds. See \cite{newell_wave_2011,newell_wave_2001} for a review.

For an infinite system, in which wave modes are indexed by a continuous wave-vector, the theory of weakly nonlinear wave turbulence becomes asymptotically exact in the weakly nonlinear limit. For finite sized systems, in which the wave modes are indexed by a discrete wave-vector, some subtleties arise. The simplest case, which is particularly relevant to numerical studies of wave turbulence, is a bi-periodic box. In this case, $\k$, is restricted to a
periodic lattice with a minimum spacing, $\Delta k$, between modes. Modulo this spacing, the components of $\k$ must be integer valued. This is an issue because if the components of $\k$ are integers, then the resonance conditions, Eq.\eqref{eq-resonance}, become a problem of Diophantine analysis. Such problems typically have far fewer solutions than their real-valued counterparts and it is generally quite difficult to find them. A complete enumeration of all solutions for the case of Eq.\eqref{eq-CHMdispersion} with $F=0$ was recently provided in \cite{bustamante_complete_2012}. This sparseness of solutions means that, in discrete systems, resonant triads can exist in isolation or in finite groups of triads known as resonant clusters. Two triads belong to the same cluster if they share at least one mode. The dynamics of small clusters consisting of two triads has been studied in considerable detail in \cite{bustamante_dynamics_2009}. Small clusters have attracted some interest in the context of atmospheric 
dynamics as a possible
explanation of the unusual periods of certain observed atmospheric oscillations \cite{kartashova_model_2007}. Depending on the dispersion relation, there may or may not exist large clusters capable of distributing energy over a
large range of scales in a discrete system. In the case of Rossby waves on a sphere with infinite deformation radius, such a large cluster does exist \cite{lvov_finite-dimensional_2009}. On the other hand, for capillary waves there are no exactly resonant triads at all \cite{kartashova_wave_1998}. For the dispersion relation Eq.\eqref{eq-CHMdispersion}, numerical explorations indicate that for general values of $F$ large exactly resonant clusters are rare. Thus, in discrete systems, it is often necessary to rely on approximate resonances to account for energy transfer.

Approximate resonance is possible due to the phenomenon known as nonlinear resonance broadening. This is an effect whereby the frequency of a wave acquires a correction to its linear value which depends on the amplitude (see Chaps. 14 \& 15 of \cite{whitham_linear_1974}). Triads which are not exactly in resonance can then interact at finite amplitude if the frequency mismatch is less than this correction.  Such triads are known as quasi-resonant triads and satisfy the broadened resonance conditions
\begin{equation}
\label{eq-quasiresonance}
\left\{
\begin{aligned}
&\k_3 = \k_1 + \k_2,\\
&\left|\omega(\k_3)-\omega(\k_1)-\omega(\k_2)\right| \leq \delta
\end{aligned}
\right.
\end{equation}
where $\delta$ is a characteristic value for the resonance broadening taken to be positive. Although Eq.\eqref{eq-quasiresonance} provides only a kinematic description of resonance broadening, the analogous dynamical effect can be very strikingly visualised in linear stability analyses of weakly nonlinear waves \cite{dyachenko_decay_2003,connaughton_modulational_2010} where it is found that the set of unstable perturbations lie in a neighbourhood around the set of exactly resonant perturbations. This set of quasi-resonant modes is often pictured as a ``thickened'' or broadened version of the exactly resonant manifold. For weakly nonlinear systems, this broadening is expected to be small since amplitudes are small. It may nevertheless be large enough to overcome frequency mismatches which arise when wave-vectors are restricted to a discrete grid and prevent discreteness from impeding the cascade of energy. A striking example of this effect is observed for capillary wave turbulence in a biperiodic box.  For 
this system, since there are no exact
resonances, as the nonlinearity is decreased the resonance broadening eventually becomes smaller than the frequency mismatches due to the grid. Direct numerical simulations illustrate that the cascade of energy to small scales stops entirely when the level of nonlinearity gets sufficiently small leading to the phenomenon of ``frozen turbulence'' \cite{pushkarev_kolmogorov_1999,pushkarev_weak_2000,connaughton_discreteness_2001}.

The interplay between exactly resonant and quasi-resonant clusters means that wave turbulence in discrete systems is nowadays believed to exhibit several regimes. If the typical resonance broadening, $\delta$, is small enough that effectively only exactly resonant clusters can interact, the dynamics are referred to as discrete wave turbulence \cite{lvov_finite-dimensional_2009,kartashova_discrete_2009,kartashova_towards_2010}. If $\delta$ is larger than the typical spacing between modes then effectively all triads can interact at least quasi-resonantly and the classical statistical theory is expected to be valid. In between is a regime consisting of a mixture of exactly resonant and quasi-resonant clusters which has been termed mesoscopic wave turbulence \cite{zakharov_mesoscopic_2005,kartashova_towards_2010}. In this intermediate regime, it has been suggested \cite{nazarenko_sandpile_2006} that forced systems could exhibit some aspects of self-organised criticality. This suggestion is motivated by the idea 
that the forcing will cause the characteristic value of $\delta$ to increase until it is large enough for a large quasi-resonant
cluster to form which will then facilitate an ``avalanche'' of energy transfer to the dissipation scale thereby reducing wave amplitudes and the corresponding value of $\delta$.

In this paper we develop a kinematic concept of criticality in quasi-resonant interactions in the Eq.\eqref{eq-CHM}. Specifically, we address the question of how a large quasi-resonant cluster emerges in the CHM equation as $\delta$ is increased. Inspired by the theory of percolation on random networks \cite{stauffer_introduction_1994}, we take ``large cluster'' to mean a cluster that consists of a finite fraction of all modes in the system. We begin by analytically characterising the shape of the quasi-resonant set defined by Eq.\eqref{eq-quasiresonance} as a function of $\delta$ for a single triad in Sec. \ref{sec-singleTriad}. By expressing the boundary of the quasi-resonant set in terms of the intersection of a pair of quadratic forms, we find some surprises. In particular, we find that the area of the set diverges at a finite value of $\delta$ illustrating that the common perception of the quasi-resonant set as a ``thickened'' version of the exact resonant manifold is potentially quite misleading. In 
Sec. \ref{sec-manyTriads} we numerically construct the set of quasi-resonant clusters as a function of $\delta$ for various system sizes. We show that a percolation transition occurs at a critical value, $\delta_*$, of the resonance broadening as $\delta$ is increased. At this critical value, the size of the largest cluster rapidly goes from containing a negligible fraction of the modes in the system to containing a finite fraction of them. The value of $\delta_*$ decreases as the inverse cube of the system size, a fact which we trace to quasi-resonant interactions between small scale meridional modes and large scale zonal modes. We finish with a short summary and discussion about what conclusions can be drawn about Rossby wave turbulence from our results.

\section{Characterisation of the quasi-resonant set for a single triad}
\label{sec-singleTriad}

In what follows, we shall take $\k_3$ to be fixed with $\k_2=\k_3-\k_1$. The $\delta$-detuned quasi-resonant set of $\k_3$ is the set of modes, $\k_1$, which satisfy the inequality
\begin{equation}
\label{eq-quasiresonance2}
\left|\omega(\k_3)-\omega(\k_1)-\omega(\k_3-\k_1)\right| \leq \delta.
\end{equation}
This section is devoted to determining the structure of this set as a function of the detuning, $\delta$. The boundaries of this set are given by the pair of curves
\begin{eqnarray}
\label{eq-plusBoundary}\omega(\k_3)-\omega(\k_1)-\omega(\k_3-\k_1) &=& \delta\,,\\
\label{eq-minusBoundary}\omega(\k_3)-\omega(\k_1)-\omega(\k_3-\k_1) &=& -\delta.
\end{eqnarray}
We begin by finding these curves. Clearly it suffices to solve Eq.\eqref{eq-plusBoundary} since the second boundary can  be obtained from this by setting $\delta \to -\delta$. To fix notation, let us write
\begin{eqnarray*}
\k_3 &=& (p,q)\,,\\
\k_1 &=& (r,s)\,,\\
k^2 &=& p^2 + q^2.
\end{eqnarray*}
For the CHM dispersion relation, Eq.\eqref{eq-CHMdispersion}, the boundary of the quasi-resonant set, Eq.\eqref{eq-plusBoundary}, then corresponds to the curve in the $(x,y)$ plane implicitly defined by
\begin{displaymath}
-\frac{\beta\, p}{k^2 + F} + \frac{\beta\, r}{r^2 + s^2 + F} + \frac{\beta\, (p-r)}{(p-r)^2 + (q-s)^2 + F} = \delta.
\end{displaymath}
Subsequent formulae will be more compact if we shift the origin to the centre of symmetry of the curve, by defining $x = r-p/2$,
$y = s  - q/2.$ Also, we rescale $\delta$ by setting $\beta = 1$ from here on. The curve we wish to study is therefore
\begin{eqnarray}
\nonumber -\frac{p}{k^2 + F} &+& \frac{x + p/2}{(x+p/2)^2 + (y+q/2)^2 + F}\\
\label{eq-curve1} &-& \frac{x - p/2}{(x-p/2)^2 + (y-q/2)^2 + F} = \delta.
\end{eqnarray}
Gathering these terms together with a common denominator we obtain the quartic curve $c(x,y)=0$, where
\begin{equation}
\label{eq-curve2}
c(x,y) = a_1 + a_2(x^2+y^2)^2 + a_3 x^2 + a_4 y^2 - a_5 x y,
\end{equation}
and the coefficients are given by
\begin{eqnarray*}
a_1 &=& \frac{1}{16}\,\left(k^2 +  4 F \right)\left[3k^2 p -\left(k^2+F\right)\left(k^2 + 4F\right)\,\delta \right]\,,\\
a_2 &=& -p - \left(k^2 +F\right)\,\delta\,,\\
%a_3 &=& -\frac{1}{2} \left[p^3 + 3 p\left(q^2 + 2 F\right) - \left[ p^2(p^2+3 F) \right. \right.\\
%& & \hspace{3.0cm} \left. \left. -(q^2+F)(q^2+4 F)\right]\,\delta \right]\\
a_3 &=& -\frac{1}{2}\left[p(p^2 + 3 q^2 + 6 F) - (k^2+F)(p^2-q^2 - 4F)\,\delta \right]\,,\\
a_4 &=& \frac{1}{2}\left[p(p^2 + 3 q^2 -2 F) - (k^2+F)(p^2-q^2+4F)\,\delta \right]\,,\\
a_5 &=& 2 q\left[ q^2 + F - p(k^2+F)\, \delta\right].
\end{eqnarray*}
If we introduce new variables $u = x^2$, $v = y^2$ and $w = x y$, then it becomes clear that the boundary curve, Eq.\eqref{eq-curve1},
corresponds to the intersection of 2 quadratic surfaces
\begin{eqnarray}
\label{eq-quadraticForms} a_2(u+v)^2 + a_3 u + a_4 v + a_1 &=& a_5 w\,,\\
\nonumber w^2 &=& u v\,.
\end{eqnarray}
Notice that the surface $w^2=u v$ is a cone, which is singular at the origin $x=y=0.$ We have performed a full analysis of the properties of the intersection curves for general values of the parameters $p$, $q$, $F$ and $\delta$. This is
a technical exercise which is not very illuminating. In the interests of clarity, we will restrict ourselves here to illustrating the essential qualitative features of these curves accompanied by some explicit examples.

\begin{figure}
\centering
\includegraphics[width=\figwidth]{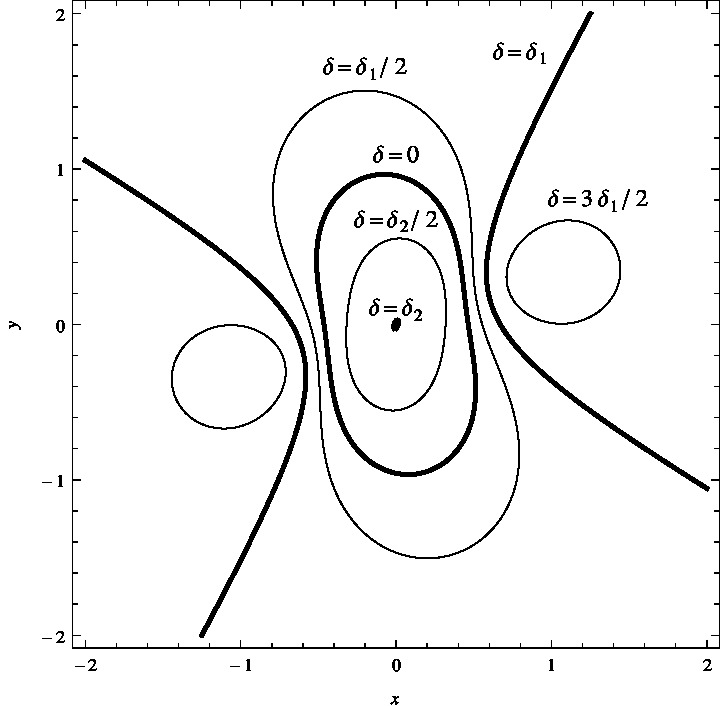}
\caption{Solutions of Eq.\eqref{eq-plusBoundary} for different values of $\delta$ with no self-intersection. Here $\k_3 =(\cos \theta, \sin \theta)$ with $\theta=\pi/6$ and $F=1/5$. The values of $\delta_1 \approx -0.721688$ and $\delta_2 \approx 1.20281$ are obtained from Eqs.\eqref{eq-delta1} and \eqref{eq-delta2} respectively.}
\label{fig-contoursNoIntersection}
\end{figure}

\begin{figure}
\centering
\includegraphics[width=\figwidth]{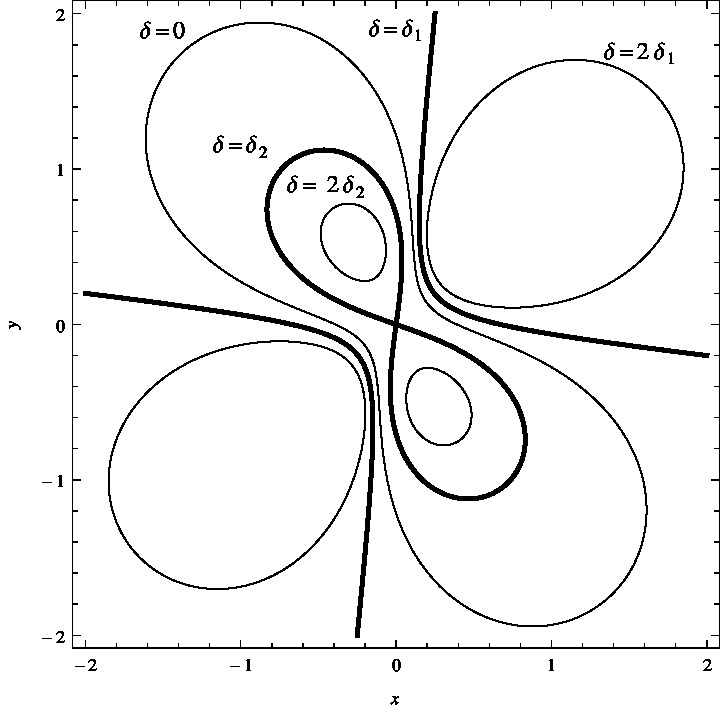}
\caption{Solutions of Eq.\eqref{eq-plusBoundary} for different values of $\delta$ which self-intersect at $\delta=\delta_2$. Here $\k_3 =(\cos \theta, \sin \theta)$ with $\theta=3\pi/7$ and $F=1/5$. The values of $\delta_1 \approx -0.185434$ and $\delta_2 \approx 0.309057$ are obtained from Eq.\eqref{eq-delta1} and Eq.\eqref{eq-delta2} respectively.}
\label{fig-contoursWithIntersection}
\end{figure}

\subsection{The curve typically exists only for a finite range of $\delta$.\\}
This is evident from studying Eq.\eqref{eq-curve1} as $\delta \to \pm \infty$. The LHS can only diverge if one of the denominators diverges. This is impossible if $F>0$. Hence if $F>0$, solutions to Eq.\eqref{eq-curve1} must cease to exist if $\delta$ gets too large by absolute value. In the exceptional case $F=0$, solutions indeed exist for all values of $\delta$ and become localised in the neighbourhood of $\pm (p/2, q/2)$ as $\delta \to \pm \infty$.

\subsection{The curve is bounded except at single critical value of $\delta$.\\}
Examining Eq.\eqref{eq-quadraticForms}, it is clear that since $u$ and $v$ are both positive, it is generally not possible to balance the quadratic terms if $u^2 + v^2 \to \infty$. Hence the curve, if it exists is bounded. The single exception to this occurs when the coefficient of the quadratic terms vanishes. The curve may therefore diverge at the single critical value of detuning given by:
\begin{equation}
\label{eq-delta1}
\delta = \delta_1 \equiv -\frac{p}{k^2 + F}.
\end{equation}
Note that this corresponds to $\delta=\w(\k_3)$. At $\delta=\delta_1$, some algebra shows that the curve is given by the hyperbola
\begin{equation}
x^2 + 2\,\frac{q}{p} \, x y - y^2 = \frac{1}{4}\,(k^2 + F).
\end{equation}
It follows that the area of the quasi-resonant set diverges as $\delta$ tends to $\delta_1$ from below. This result illustrates that the common conception that the quasi-resonant set looks like a ``thickened'' version of the exact resonant manifold is a misconception.  The special case $p\to 0$ corresponding to the case of $\k_3$ becoming zonal is discussed separately below.

\subsection{The curve may self-intersect only at a single critical value $\delta$.\\}
From Eq.\eqref{eq-quadraticForms}, self-intersection is only possible if the curve passes through $(0,0)$ in the $(u,v)$ plane. This can only happen if the coefficient $a_1 =0$. Thus we identify a second critical value of $\delta$ where self-intersection may occur:
\begin{equation}
\label{eq-delta2}
\delta = \delta_2 = \frac{3 k^2 p}{(k^2+F)(k^2 + 4 F)}.
\end{equation}
Note that $a_1=0$ does not necessarily mean the curve self-intersects. It is possible that at $\delta=\delta_2$, the curve reduces to a single
point. To probe what happens at $\delta=\delta_2$, we consider the surface $z=c(x,y)$ defined by Eq.\eqref{eq-curve2} when $\delta=\delta_2$ in the neighbourhood of the origin. Calculation of the partial derivatives indicate that this surface has a critical point at the origin. After some tedious algebra, we find that the determinant of the matrix of second derivatives is
\begin{displaymath}
\Delta = -\frac{4 (k^2+F)^2(k^4 + 4 (q^2-3p^2)F)}{k^2+4 F}.
\end{displaymath}
If $\Delta>0$ then the critical point at $(0,0)$ is a maximum or a minimum. The curve $c(x,y)=0$ is then an isolated point. On the other hand, if $\Delta<0$
the critical point is a saddle and the curve $c(x,y)=0$ has a self-intersection at $(0,0)$. The condition for self-intersection is therefore
\begin{equation}
k^4 + 4 (q^2-3p^2)F >0.
\end{equation}
We note that for $F=0$ we always have a self-intersection.

These qualitative features of the boundary of the quasi-resonant set are illustrated graphically in Figs. \ref{fig-contoursNoIntersection} and \ref{fig-contoursWithIntersection} which show the shape of the curve for different values of $\delta$. Generic parameters, specified in the captions, have been chosen with no particular symmetries. Hence the curves shown in these figures are representative of what is found from the complete analysis of Eq.\eqref{eq-curve1}.  Fig. \ref{fig-contoursNoIntersection} shows an example in which no self-intersection occurs at $\delta=\delta_2$, while Fig. \ref{fig-contoursWithIntersection} shows an example in which a self-intersection occurs. The hyperbolic curves identified at $\delta=\delta_1$ are clearly visible in both cases.

\begin{figure*}
\centering
\subfigure[ $\delta=0$]{
\label{fig-p1q0delta0}
\includegraphics[width=0.6\figwidth]{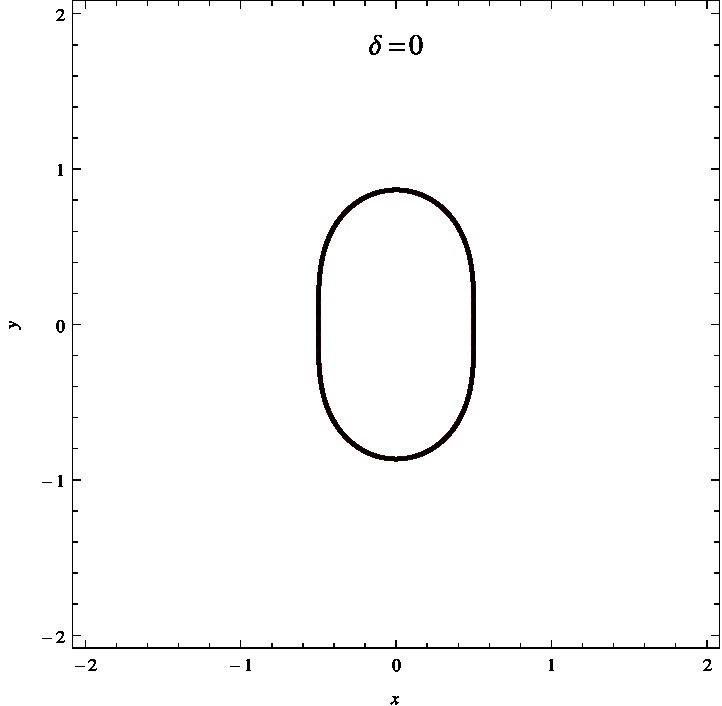}
}
\subfigure[$\delta=1/2$]{
\label{fig-p1q0delta0p5}
\includegraphics[width=0.6\figwidth]{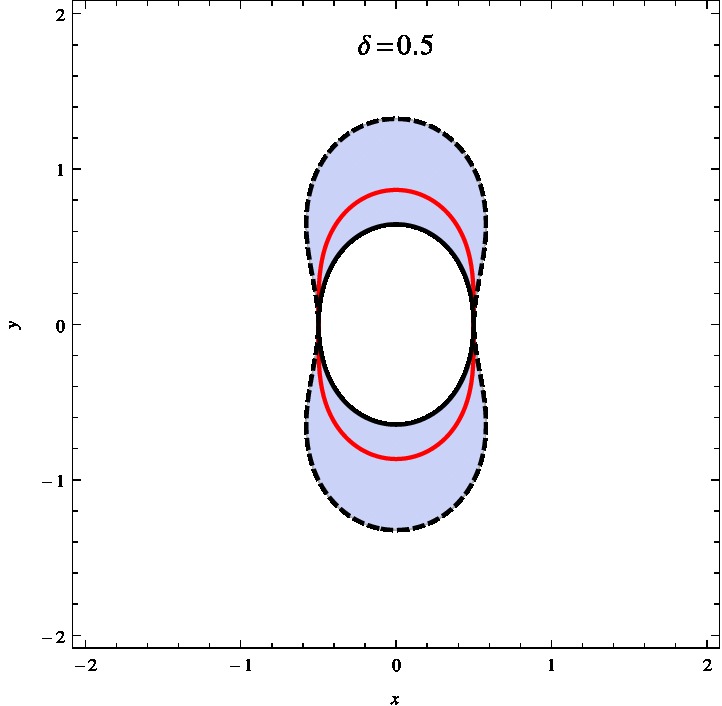}
}
\subfigure[$\delta=1$]{
\label{fig-p1q0delta1}
\includegraphics[width=0.6\figwidth]{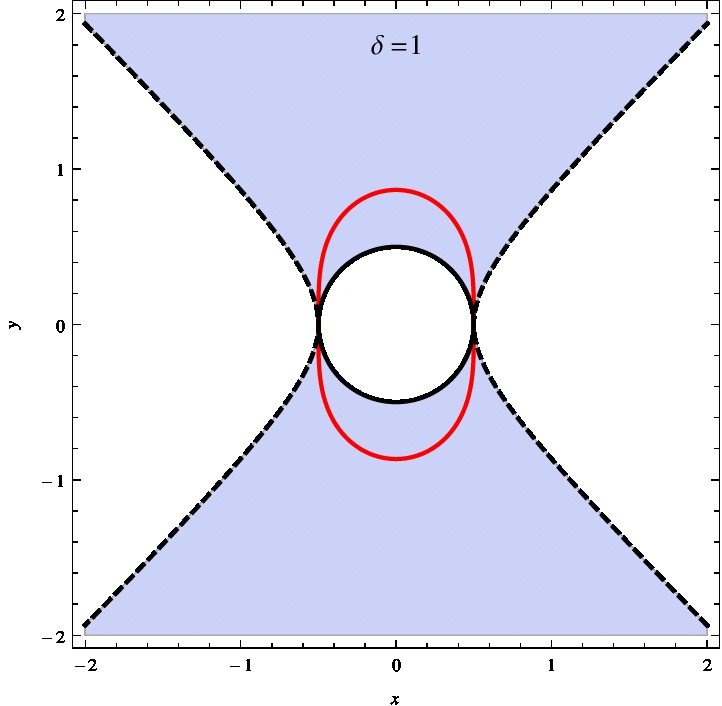}
}
\subfigure[$\delta=2$]{
\label{fig-p1q0delta2}
\includegraphics[width=0.6\figwidth]{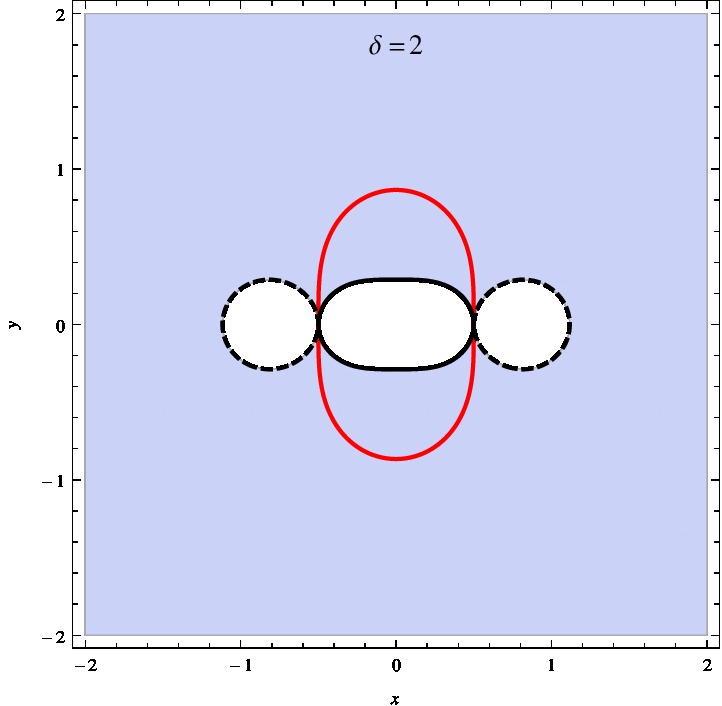}
}
\subfigure[$\delta=3$]{
\label{fig-p1q0delta3}
\includegraphics[width=0.6\figwidth]{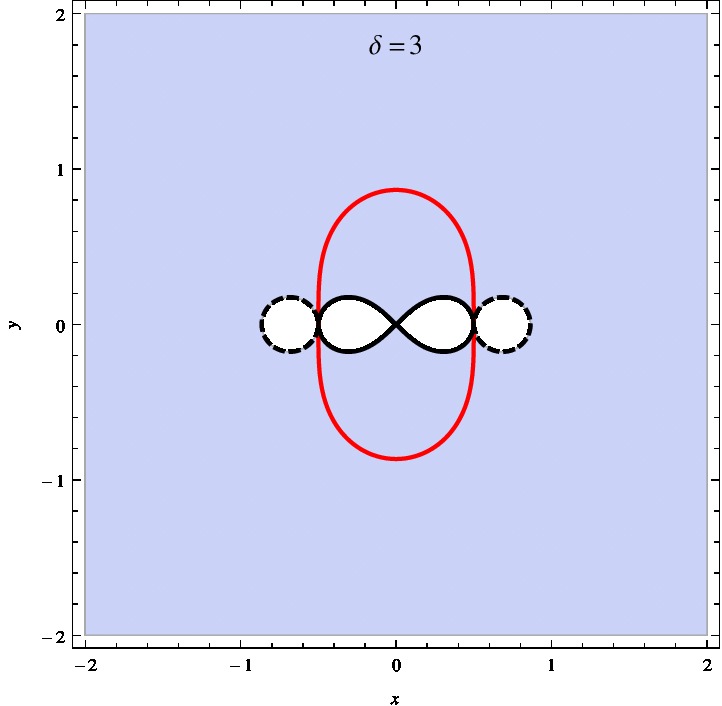}
}
\subfigure[$\delta=4$]{
\label{fig-p1q0delta4}
\includegraphics[width=0.6\figwidth]{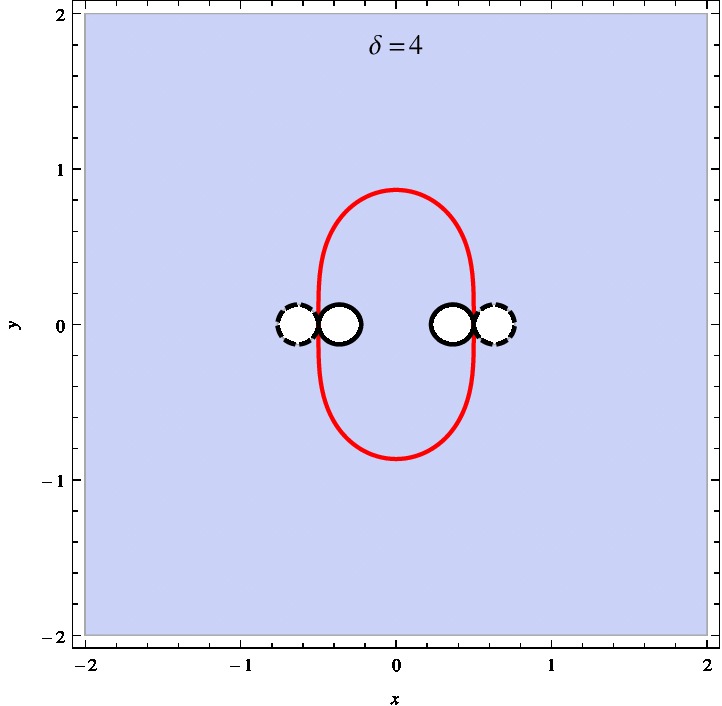}
}
\caption{\label{fig-meridionalSet} Shaded regions correspond to the resonant set defined by Eq.\eqref{eq-quasiresonance} with $\k_3=(1,0)$, $F=0$ and $\beta=1$ for different values of $\delta$. The exactly resonant manifold is the red solid line. }
\end{figure*}

\begin{figure*}
\centering
\subfigure[ $\delta=0$]{
\label{fig-p0q1delta0}
\includegraphics[width=0.6\figwidth]{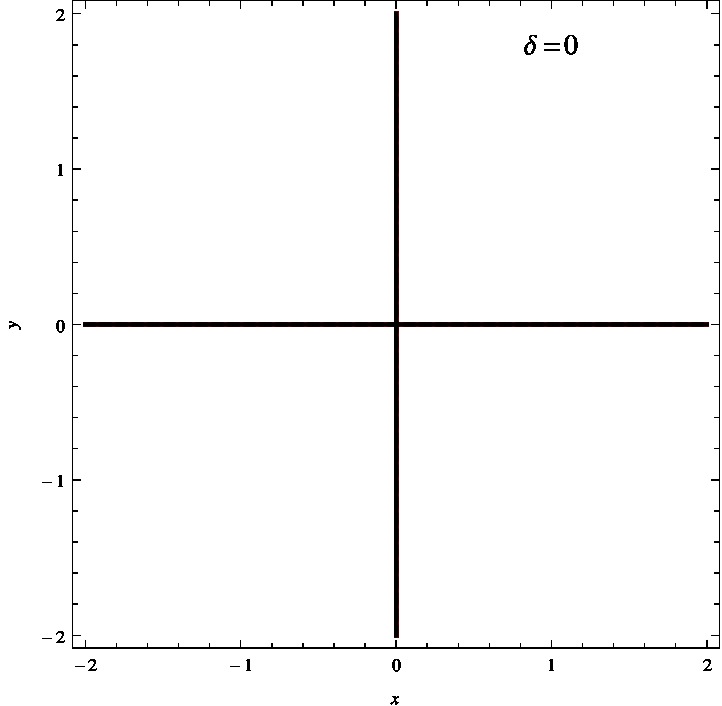}
}
\subfigure[$\delta=1/10$]{
\label{fig-p0q1delta0p1}
\includegraphics[width=0.6\figwidth]{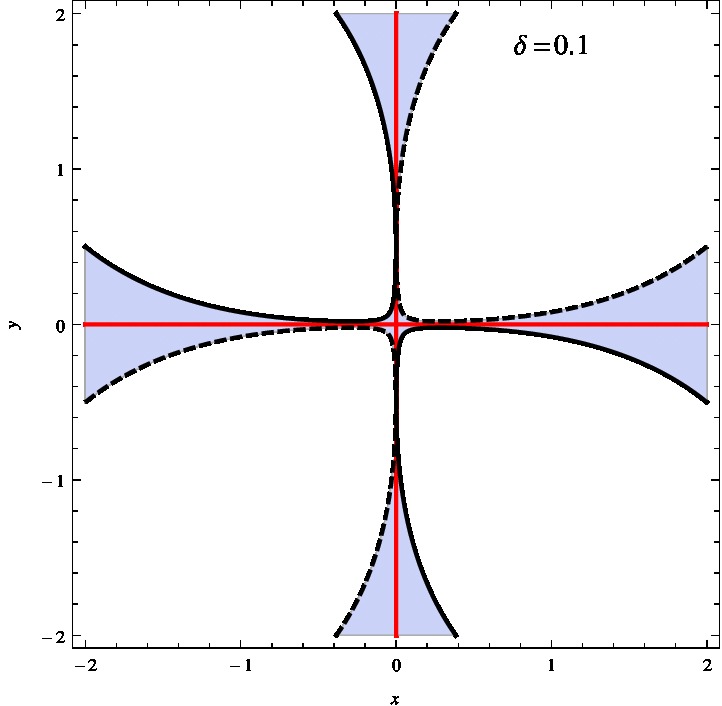}
}
\subfigure[$\delta=1/2$]{
\label{fig-p0q1delta0p5}
\includegraphics[width=0.6\figwidth]{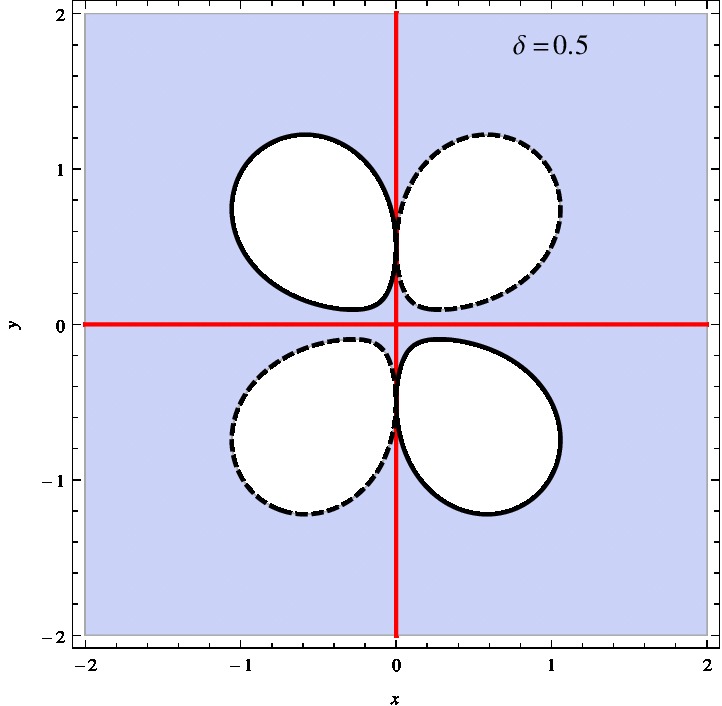}
}
%\subfigure[$\delta=2$]{
%\label{fig-p0q1delta2}
%\includegraphics[width=\figwidth]{./p0q1delta2.eps}
%}
\caption{\label{fig-zonalSet}Shaded regions correspond to the resonant set defined by Eq.\eqref{eq-quasiresonance} with $\k_3=(0,1)$, $F=0$ and $\beta=1$ for different values of $\delta$. The exactly resonant manifold is the red solid line and corresponds to the two axes in this case. As explained in the remark below equation (\ref{eq-zonalCase}), this exactly resonant manifold consists of non-generic triads, i.e., triads where one or more interaction coefficients are identically zero. The case $F=0$ is chosen for simplicity only: for $F>0,$ the resonant sets have the same qualitative shape as in the case $F=0.$ }
\end{figure*}

A couple of special cases are worth noting:
\begin{itemize}
\item[(i)] {$p=1$, $q=0$ and $F=0$\\}
This corresponds to a meridional mode. In this case, $a_5=0$ so the intersection of quadratic forms given by Eq.\eqref{eq-quadraticForms} lies entirely in the $w=0$ plane and reduces to
\begin{equation}
\label{eq-meridionalCase}
-(1 + \delta) (u+v)^2 - \frac{1}{2}(1-\delta)(u-v) + \frac{1}{16}(3-\delta) = 0.
\end{equation}
We can immediately identify the critical points. The curve self-intersects at $\delta=3$ and diverges at $\delta=-1$. The point $\delta=1$ is also noteworthy, being the complementary boundary to the divergent case. For this value of $\delta$ the curve is a perfect circle.
\item[(ii)] {$p=0$, $q=1$ and $F=0$\\}
This corresponds to a zonal mode. In this case the curve simplifies to
\begin{equation}
\label{eq-zonalCase}
\delta\, (u+v)^2 + \frac{1}{2}\,\delta\,(u-v) + \frac{1}{16}\,\delta = - 2\,w.
\end{equation}
The only special value of $\delta$ for this case is $\delta=0$. We then recover the exact resonant manifold of a zonal mode, $x\, y =0$. This consists of the two coordinate axes. It is now less surprising that the boundary of the quasi-resonant set can diverge for finite  $\delta$ once we appreciate that the exact resonant manifold of a zonal mode is unbounded. The divergence of the boundary of the quasi-resonant set of the non-zonal modes in some sense reflects the presence of this structure in the dispersion relation.

\end{itemize}

\noindent \textbf{Remark.} Notice that the exactly resonant manifold in case (ii) consists of non-generic triads, i.e., triads where one or more interaction coefficients are identically zero. A point on the $x$-axis corresponds to a so-called catalytic interaction ($|\k_1| = |\k_2|$), for which the zonal mode $\k_3$ does not change its energy but influences the energy exchange between the other two modes in the triad. A point on the $y$-axis corresponds to a `spurious' triad, formed by purely zonal modes, which do not interact at all: the interaction coefficients are all identically zero because the three modes are collinear. In our computations of the network of quasi-resonant modes (Section \ref{sec-manyTriads}), non-generic triads are discarded from the start.

Having given a fairly complete qualitative description of the behaviour of the curves which define the boundaries of the resonant sets as the detuning is varied, it now remains to assemble these boundary curves for postive and negative values of $\delta$ to determine the interior and exterior of the quasi-resonant set. This is again best accomplished by illustration. Figs. \ref{fig-meridionalSet} and \ref{fig-zonalSet} illustrate the quasi-resonant set for the two special cases discussed above for a range of increasing values of $\delta$. The shaded areas in these figures correspond to the quasi-resonant sets. The exact resonant curve is also shown for reference. Fig.\ref{fig-meridionalSet} illustrates one of the key points of this article: the common picture of the quasi-resonant set as a thickened version of the exact resonant manifold is appropriate only for small values of the broadening (eg Fig.\ref{fig-p1q0delta0p5}). One might counter this with the observation that it is only in the weakly nonlinear
regime that it makes sense to be discussing quasi-resonant interactions in the first place and in this regime, the broadening is necessarily small. Eq.\eqref{eq-delta1} tells us, however, that no matter how small the broadening, there are always modes close to the zonal axis, whose quasi-resonant set diverges.

\section{Structure of the network of quasi-resonant modes}
\label{sec-manyTriads}

\begin{figure}
\centering
\includegraphics[width=\figwidth]{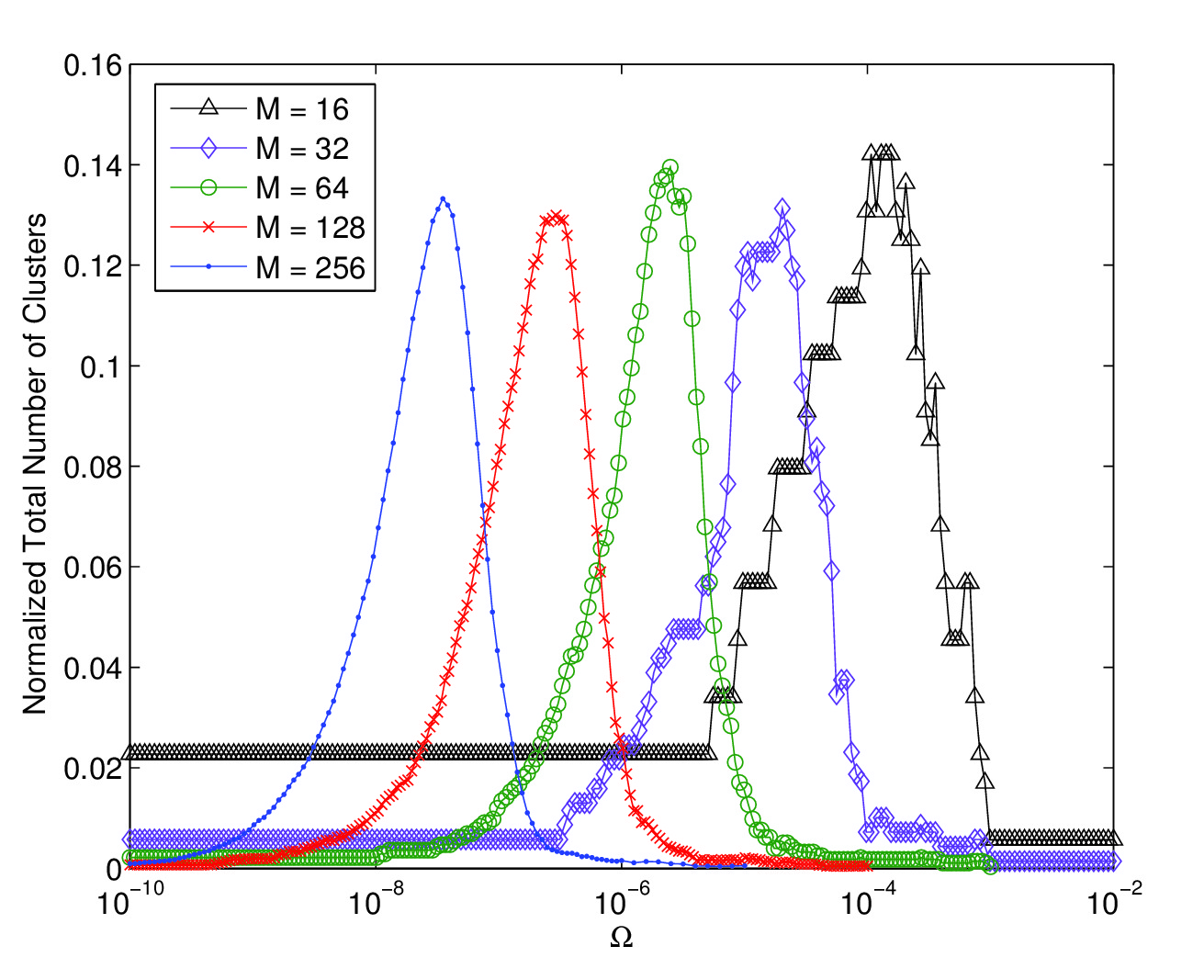}
\caption{The normalised total number of clusters, $N$, plotted as a function of $\delta$ for different system sizes, $M$. Here, we have fixed $F=\beta=1$.}
\label{fig-CHM_total_numbers}
\end{figure}

\begin{figure}
\centering
\includegraphics[width=\figwidth]{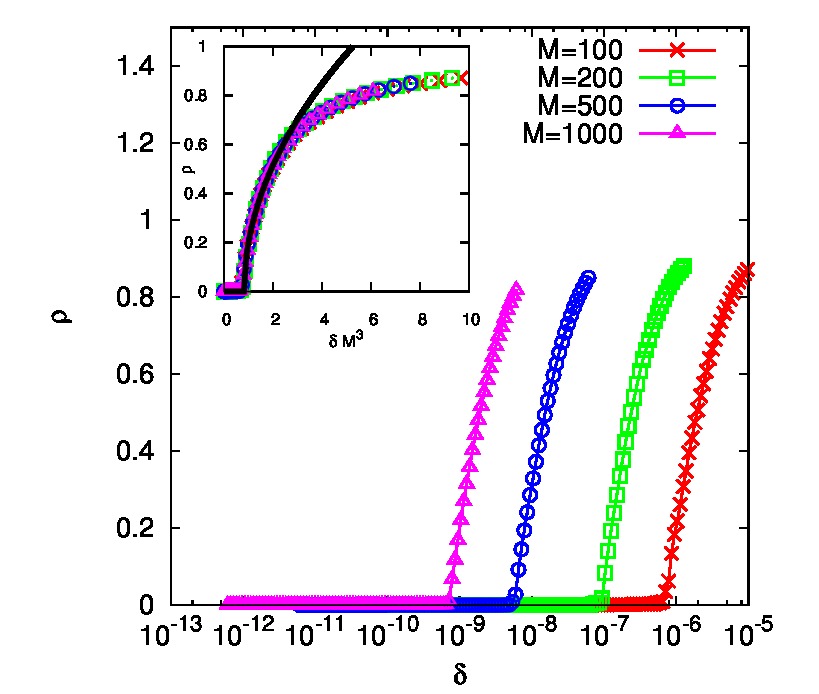}
\caption{The density of the largest cluster, $\rho$, as a function of resonance broadening $\delta$ for different system sizes, $M$. Here we have fixed $F=\beta=1$. The percolation threshold, $\delta_*$, decreases as a function of system size. The inset shows the same data plotted as a function of $\delta\,M^3$. The solid line is a fit of Eq.\eqref{eq-phaseTransition} to the data.}
\label{fig-CHM_density}
\end{figure}

In a turbulent system many modes are excited. The fact that a mode can be a member of more than one quasi-resonant set leads to overlap between sets.  This allows modes to join together to form a network of quasi-resonant clusters analogous to the exactly resonant clusters which have been extensively studied in the literature to date. In this section we study the structure of this quasi-resonant network as the typical amount of broadening, $\delta$, in the system is varied. We consider a finite system of size $2\pi \times 2\pi$ and truncate the wavenumber space such that there are $M$ modes in the $k_x$ and $k_y$ directions, $M^2$ modes in all. The spacing between modes is $2\pi/M$. The quasi-resonant clusters were obtained by an exhaustive numerical search speeded up by incorporating symmetries of the triads. As explained in the remark below equation (\ref{eq-zonalCase}), we have eliminated from our list of triads the so-called non-generic triads, i.e., the triads for which one or more interaction 
coefficients are zero. This includes triads formed out of collinear modes and triads where any two wave-vectors have the same modulus.

We first measure the total number of clusters normalised by the maximum number of possible isolated clusters which is approximately $M^4/3$. This is shown as a function of $\delta$ for several different system sizes, $M$, in Fig. \ref{fig-CHM_total_numbers}. For each system size, as $\delta$ increases, the total number of clusters first increases, then reaches a maximum at a particular value of $\delta$, which we shall denote by $\delta_*$, and then decreases.
The value, $\delta_*$, at which the maximum occurs decreases as the system size, $M$, is increased. In order to understand these results one must realise that for the CHM dispersion relation, the number of exactly resonant clusters which live on the grid of discrete wavevectors is rather small with most discrete triads necessarily exhibiting a small amount of detuning enforced by the geometry. The initial growth in the number of clusters is explained by the addition of isolated triads to the list of quasi-resonant clusters as the broadening grows large enough to incorporate the geometrical detuning generated by the grid. While one would expect this rate of increase to slow down as clusters start joining to form larger clusters, the attainment of a maximum and subsequent decrease in the total number of clusters becomes much easier to understand when we measure the size of largest cluster, $\rho$, which is simply the number of modes in the largest cluster normalised by the system size, $M^2$. This is shown in
Fig. \ref{fig-CHM_density}. We see that as $\delta$
approaches $\delta_*$ the largest cluster quickly makes a transition from including a negligible fraction of the total number of modes in the system to including almost all of them. Thus the network of quasi-resonant modes undergoes a percolation transition at $\delta=\delta_*$. While the results shown in Figs. \ref{fig-CHM_total_numbers} and \ref{fig-CHM_density} were obtained for $\beta=F =1$, we checked that the results are not very sensitive to this choice. Although this is an entirely kinematic study, we expect that the emergence of a percolating quasi-resonant cluster is the explanation for the dynamical transition from discrete wave turbulence to Kolmogorov turbulence in the CHM equation as the nonlinearity of the system is increased.

\begin{figure}
\centering
\includegraphics[width=\figwidth]{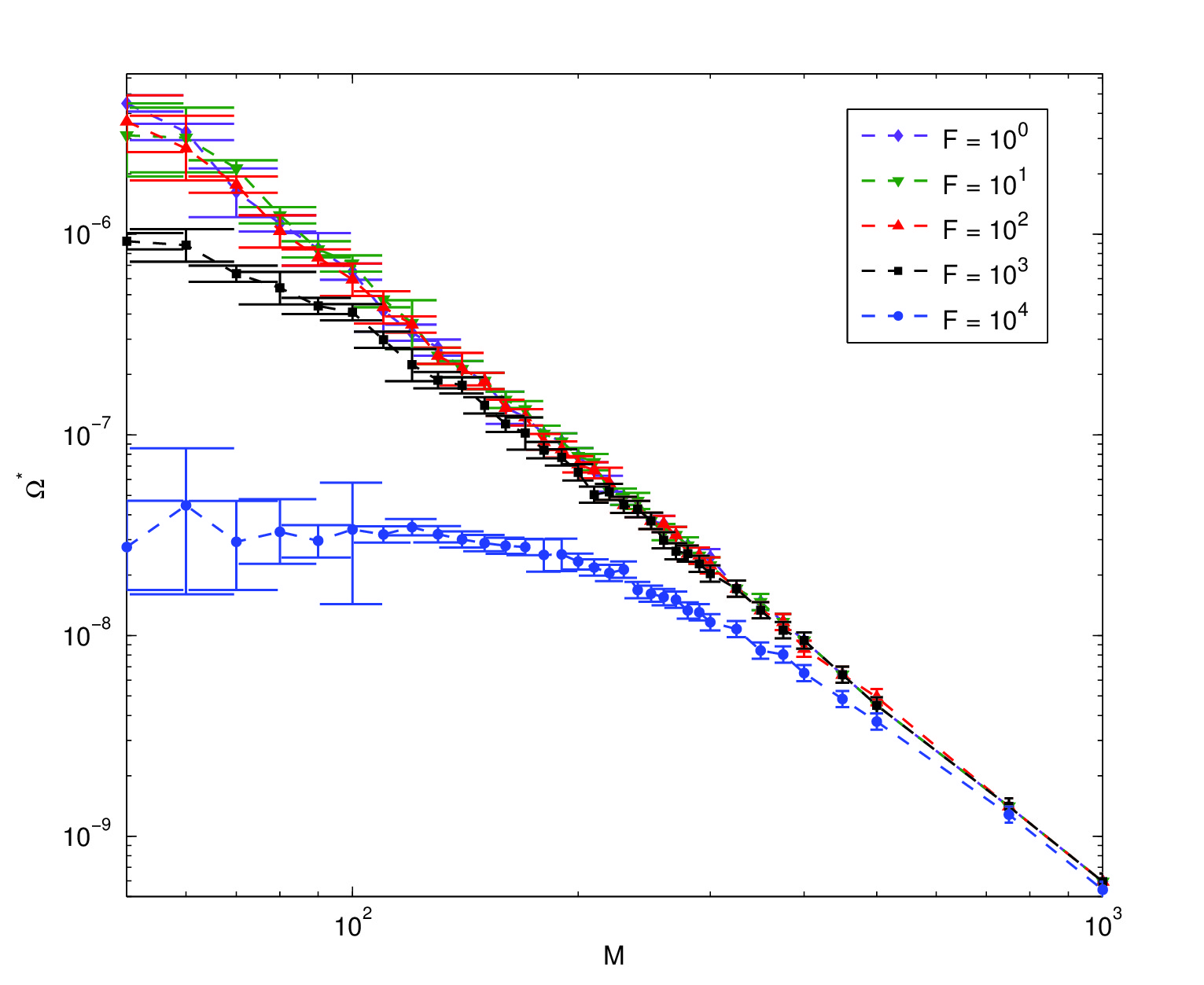}
\caption{Percolation threshold, $\delta_*$, plotted against system size and for different values of $F$. Here, we have fixed $\beta=1$. For large system sizes, $\delta_*$ exhibits the scaling $\delta_* \sim M^{-\sigma_1}$, where we measure $\sigma_1 = 3.00\;(2.97,3.03)$. The brackets show the 95\% confidence interval.}
\label{fig-critical_points}
\end{figure}

The decrease of the percolation threshold, $\delta_*$, as the system size, $M$, increases is clearly due to the fact that as $M$ increases, the spacing
between discrete modes decreases. The amount of broadening required to overcome the geometrical detuning is therefore smaller and hence clusters can
form more easily. We would like to quantify this decrease. We have already learned that the area of the quasiresonant set of any particular mode, $\k$, diverges at a finite value of $\delta=\w(\k)$ as indicated in Eq.\eqref{eq-delta1}. It seems plausible that as soon as the broadening is sufficiently large to allow any mode in the system to have a divergent quasi-resonant set, then the largest cluster must be of the size of the system. Thus one might estimate
\begin{equation}
\label{delta_sat1}
\delta_*\approx\min_{\k}|\omega(\k)|.
\end{equation}
By this argument, Eq.\eqref{eq-CHMdispersion} tells us that for large systems we should see the scaling, $\delta_* \sim M^{-2}$.  A numerical measurement of how $\delta_*$ scales with system size for a range of values of $F$, is plotted in figure \ref{fig-critical_points}. We see that the value of $F$ becomes redundant for large $M$ as already mentioned. Each of the curves converge to the straight line indicated by the scaling
\begin{equation}
\delta_*\sim M^{-\sigma_1},
\end{equation}
with exponent $\sigma_1 = 3.00$ with a 95\% confidence intervals if $(2.97,3.03)$, which was obtained by bias-corrected bootstrapping. The connected component therefore forms much more easily than the naive argument above would suggest. The origin of this $M^{-3}$ scaling can be traced to the
fact that zonal modes require very little detuning in order to interact with high $\k$ almost-meridional modes. Take $F=0$ for simplicity, although the following argument works for any $F>0.$ Let us consider the largest-scale
zonal mode in the system, $\k=(0,1).$ We recall that the exactly resonant manifold $x y = 0$ of this zonal mode, figure \ref{fig-zonalSet}(a), gives rise to non-generic triads which do not interact efficiently and therefore must be discarded. So we ask what is the minimum value of the detuning so that the  quasiresonant set of the mode $\k=(0,1)$ contains some new modes. From Fig. \ref{fig-zonalSet}(b) it is clear that the first new mode to join the quasi-resonant set will be $\k_1 = (M,0),$ corresponding to $x=M, y = -1/2.$ By directly replacing these values of $x,y$ into Eq.\eqref{eq-zonalCase}, we obtain
\begin{displaymath}
\left[ \left(M^2+\frac{1}{4}\right)^2 + \frac{1}{2}\left(M^2 - \frac{1}{4}\right) + \frac{1}{16}\right] \, \delta_* = M\,.
\end{displaymath}
For large $M$, this has solution $\delta_* \sim {M^{-3}}$, in agreement with Fig. \ref{fig-critical_points} and the inset of Fig. \ref{fig-CHM_density}. This suggests that the percolation transition is driven by interactions between large scale zonal modes and small scale meridional modes. This is consistent with the known scale-nonlocality of wave turbulence in the CHM equation (see \cite{connaughton_feedback_2011} and the references therein) and provides further evidence, albeit circumstantial, that the percolation transition is associated with the onset of turbulence in the CHM model.

\begin{figure}
\centering
\includegraphics[width=\figwidth]{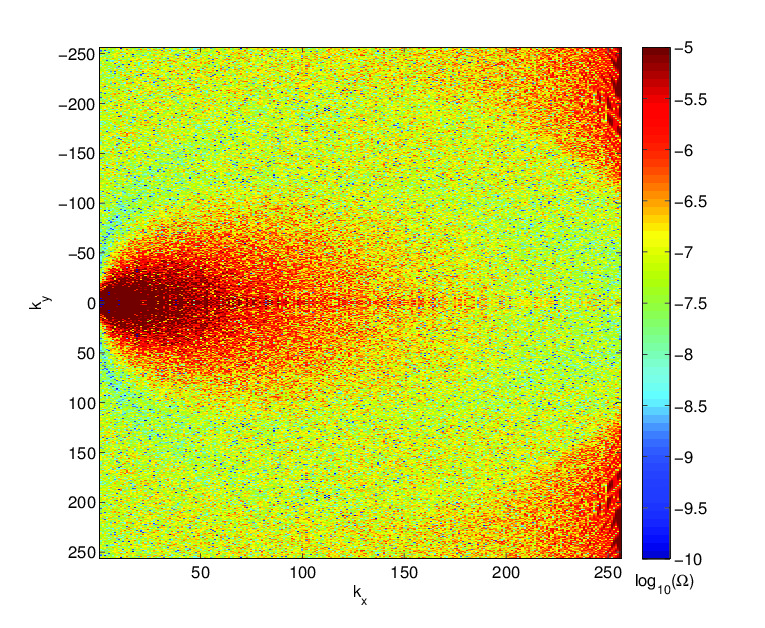}
\caption{Color map in $\k$-space showing the  smallest value of broadening required for each mode to become a member of a cluster of any size. The darker regions are resilient to becoming members of clusters  in the sense that large values of broadening are required. Lighter regions join clusters easily.}
\label{fig-M256_missing_nodes}
\end{figure}

In order to further illustrate this point, we can ask which modes have the greatest or least tendency to become part of a quasi-resonant cluster. This is done in figure (\ref{fig-M256_missing_nodes}), where we have coloured each mode according to the minimal amount of resonance broadening required for this mode to join a cluster of any size. Blue modes become active very easily whereas red modes are resilient to becoming part of any cluster. Rather than appearing homogeneous and random, we see the appearance of definite structure. We see a circular region containing modes with a strong propensity to join clusters including a narrow large-scale region of zonal modes accumulating near the $k_x=0$ axis with very low interaction thresholds as expected from the discussion above.  We also remark upon the group of large-scale  meridional scales reluctant to form any quasi-resonant connections. The relative reluctance of large-scale meridional modes to exchange energy has already been remarked up in the literature
and suggested as an explanation of the inherent anisotropy of Rossby waves turbulence. For example in \cite{vallis_atmospheric_2006}, a wave-turbulence boundary was computed by comparing the CHM dispersion relation to the inverse of the eddy-turnover time. It is then argued that inside this region Rossby waves dominate, but with a frequency incommensurate with that of the surrounding turbulence so that energy cannot penetrate into this region. The boundary thus obtained  seems similar in position and shape to the dark region in Fig. \ref{fig-M256_missing_nodes}. This anisotropic energy distribution at large scales has been documented in numerical simulations of Rossby wave turbulence (see \cite{huang_anisotropic_2001} and the references therein). It is somewhat surprising to see it emerging again here from purely kinematic considerations.

\begin{figure}[tb]
\centering
\includegraphics[width=\figwidth]{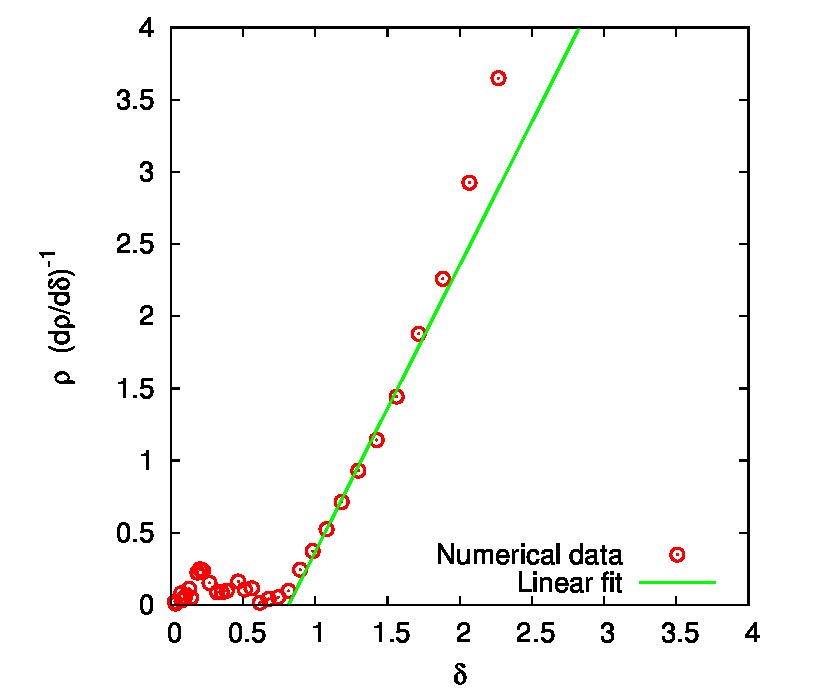}
\caption{Fitting the data in Fig.\ref{fig-CHM_density} to a standard phase transition profile. The lefthand side of Eq.\eqref{eq-phaseTransition} is plotted as a function of $\delta$. The fit (solid line) is taken over the range $\left[\delta_*:2\right]$ with $\delta_*=0.815$. It has slope 2.03. which suggests a value of $z\approx \frac{1}{2}$.}
\label{fig-fitting}
\end{figure}

Finally we might ask whether the density of the giant cluster illustrated in Fig. \ref{fig-CHM_density} shows the generic profile for a second order phase transition,
\begin{equation}
\label{eq-phaseTransition}
\rho(\delta) = \left\{ \begin{aligned}
&0 &\mbox{$\delta\leq \delta_*$}\\
&c\,(\delta-\delta_*)^z & \mbox{$\delta > \delta_*$},
\end{aligned}
\right.
\end{equation}
and, if so, what is the exponent $z.$ Notice that the system size $M$ has been absorbed after appropriate rescalings of $\delta$ and $\delta_*$ by the factor $M^3.$  Eq.\eqref{eq-phaseTransition} contains 3 adjustable parameters, $\delta_*$, $c$ and $z$, which makes it difficult to unambiguously determine the exponent $z$. As pointed out in \cite{connaughton_warm_2004}, if Eq.\eqref{eq-phaseTransition} holds, then
\begin{equation}
\rho\ \left(\dd{\rho}{\delta}\right)^{-1} = \frac{1}{z}\left( \delta-\delta_*\right)
\end{equation}
for $\delta>\delta_*$. Plotting this quantity against $\delta$ produces an easier fitting problem because the amplitude $c$, cancels out, the fitting becomes linear and the value of $\delta_*$ corresponds to the point at which the fitted line crosses zero. The values of $\dd{\rho}{\delta}$ would ideally be obtained independently from the values for $\rho$. In our case, this was not possible so we obtained them by locally interpolating the measured values of $\rho$ using $Mathematica$ and differentiating the result. The outcome of this analysis is shown in Fig. \ref{fig-fitting}. We can see that a case can be made for a linear fit in the neighbourhood of $\delta_*$. Taking $\delta_*=0.815$ (the point at which the straight line fit obviously starts to fail), and fitting the data over the range $\left[\delta_*:2\right]$ gives the fit shown in the figure. The slope gives a value of $z\approx \frac{1}{2}$. This is standard Landau value for the mean field theory of a second order phase transition of a scalar field.
 These results, while suggestive, are far from definitive. A more detailed numerical study in the vicinity of $\delta_*$ will be required before we can start putting estimates of uncertainty on these values. For the purposes of comparision, Eq.\eqref{eq-phaseTransition} with the best fit values of the parameters, is plotted with the original data in the inset of Fig.\ref{fig-CHM_density} (solid line).

\section{Conclusions and outlook}
\label{sec-conclusion}

To conclude, we have presented a kinematic analysis of the properties of quasi-resonant triads in the CHM equation. We described the analytic form of the quasi-resonant set defined by the quasi-resonance conditions, Eq. \eqref{eq-quasiresonance} as a function of the resonance broadening, $\delta$. We found that they have non-trivial geometric shape and are not well described as simply thickened versions of the exact resonant manifold as is commonly assumed. In particular, we found that the quasi-resonant set becomes unbounded for above a critical value of $\delta$ and that this can occur for arbitrarily small values of $\delta$ as we consider modes approaching the zonal axis. We then conducted an in-depth numerical study of the structure of quasi-resonant clusters as a function of $\delta$ and identified a percolation transition as $\delta$ is increased. At the transition, a large cluster is formed which contains a finite fraction of all the modes in the system. For a system containing $M^2$ modes, the value
of the percolation threshold decreases as $M^{-3}$. This scaling results from the ease with which large scale zonal modes interact with small scale meridional modes, a reflection of the nonlocality of Rossby wave turbulence.

We speculate that the percolation transition corresponds dynamically to the transition from mesoscopic to classical wave turbulence. The fact that a percolation transition exists is consistent with earlier work on capillary waves \cite{connaughton_discreteness_2001}. In fact, we believe that this transition is not a special feature of the CHM dispersion relation and is generic \cite{harris_kinematics_2013}. Furthermore our results provide circumstantial support for the sandpile picture of mesoscopic wave turbulence suggested in \cite{nazarenko_sandpile_2006} since a small change in resonance broadening in the vicinity of $\delta_*$ can trigger a transition from a state which cannot support an energy cascade to one which can. In order to better understand these issues, we believe that it is important to move beyond the kinematic picture of resonance broadening and attempt to devise methods of studying these effects dynamically.

\begin{acknowledgments}
C.C. acknowledges the support of the EPSRC grant EP/H051295/1. \, M.D.B. acknowledges support from University College Dublin, Seed Funding Projects SF564 and SF652.

\end{acknowledgments}

\bibliography{main}

\end{document}